\documentclass[letterpaper,12pt]{article}
\usepackage{mathptmx}
\usepackage{amsmath}
\usepackage{amsfonts}
\usepackage{amsthm}
\usepackage{mathtools}
\usepackage{amssymb}
\usepackage[pdftex]{graphicx}
\usepackage{stmaryrd}
\usepackage{float}
\usepackage{mathrsfs}
\usepackage{bbold}
\usepackage{caption}
\usepackage{BOONDOX-cal}
\usepackage{hyperref}

\usepackage{graphics}

\usepackage[authoryear,comma,sectionbib, sort]{natbib} 

\usepackage{subfig}

\DeclareMathAlphabet{\mathcall}{U}{BOONDOX-cal}{m}{n}
\SetMathAlphabet{\mathcall}{bold}{U}{BOONDOX-cal}{b}{n}

\usepackage{tikz}
\usetikzlibrary{arrows,shapes}
\tikzstyle{var}=[draw=none,fill=none]
\tikzstyle{unobsvar}=[circle, draw,fill=none]
\tikzstyle{edge} = [draw,thick,->]
\tikzstyle{edge2} = [draw,thick, dashed,->]  
\usetikzlibrary{arrows,shapes}
\tikzstyle{edge2gris} = [draw,thick, dashed,->,CadetBlue]
\tikzstyle{edgebleu} = [draw,thick,->, NavyBlue]
\tikzstyle{edgerouge} = [draw,thick,->, OrangeRed]
\tikzstyle{edgeBleu} = [draw,thick,dotted,->, NavyBlue]
\tikzstyle{edgeRouge} = [draw,thick, dotted, ->, OrangeRed]
\tikzstyle{edgegris} = [draw,thick, ->, gray]

\renewcommand{\P}{{\rm I}\kern-0.14em{\rm P}}
\newcommand{\p}{{\rm I}\kern-0.18em{\rm P}}
\newcommand{\E}{{\rm I}\kern-0.14em{\rm E}}

\def\1{{\rm 1\mskip-4.4mu l}}

\renewcommand{\P}{\mathbb{P}}

\usepackage{color}

\newcommand{\R}{\mathbb{R}}

\makeatletter
\newcommand*{\indep}{%
  \mathbin{%
    \mathpalette{\@indep}{}%
  }%
}
\newcommand*{\nindep}{%
  \mathbin{
    \mathpalette{\@indep}{\not}
  }%
}
\newcommand*{\@indep}[2]{%
  \sbox0{$#1\perp\m@th$}
  \sbox2{$#1=$}
  \sbox4{$#1\vcenter{}$}
  \rlap{\copy0}
  \dimen@=\dimexpr\ht2-\ht4-.2pt\relax
  \kern\dimen@
  {#2}%
  \kern\dimen@
  \copy0 
}

\newtheorem{Proposition}{Proposition}

\usepackage{ulem}

\author{Lola Étiévant*, Vivian Viallon}
\title{On some limitations of probabilistic models for dimension-reduction: Illustration in the case of probabilistic formulations of partial
least squares}
\date{}

\pdfoutput=1

\begin{document}

\maketitle

\noindent *Corresponding author: Lola Étiévant, Institut Camille Jordan, Villeurbanne 69622, France, E-mail: lola.etievant@gmail.com. https://orcid.org/0000-0001-7562-3550\\

\noindent Vivian Viallon, Nutritional Methodology and Biostatistics, International Agency for Research on Cancer, Lyon 69372, France, E-mail: viallonv@iarc.fr\\

\noindent The final published article and its Supporting Information are available at the \textcolor{red}{\href{https://onlinelibrary.wiley.com/doi/abs/10.1111/stan.12262}{Statistica Neerlandica}} website.

\begin{abstract}
Partial Least Squares (PLS) refer to a class of dimension-reduction techniques aiming at the identification of two sets of components with maximal covariance, to model the relationship between two sets of observed variables $x\in\mathbb{R}^p$ and $y\in\mathbb{R}^q$, with $p\geq 1, q\geq 1$. Probabilistic formulations have recently been proposed for several versions of the PLS. Focusing first on the probabilistic formulation of the PLS-SVD proposed by \cite{elBouhaddani17}, we establish that the constraints on their model parameters are too restrictive and define particular distributions for $(x,y)$, under which components with maximal covariance (solutions of PLS-SVD) are also necessarily of respective maximal variances (solutions of principal components analyses of $x$ and $y$, respectively). We propose an alternative probabilistic formulation of PLS-SVD, no longer restricted to these particular distributions. We then present numerical illustrations of the limitation of the original model of \cite{elBouhaddani17}. We also briefly discuss similar limitations in another latent variable model for dimension-reduction.

\textbf{keywords: }PLS; probabilistic formulation; identifiability
\end{abstract}

\section{Introduction}\label{Sec:Intro}

Principal Component Analysis (PCA), Canonical Correlation Analysis (CCA) and Partial Least Squares (PLS) are arguably among the most popular multivariate methods for dimension-reduction. They have been described and applied for many years \citep{Hotelling36, Wold82, Wold85, Hotelling33, Sampson89_PLSSVD}, and are still the subject of active research and discussion \citep{Abdi13, Jolliffe16, Jolliffe02, Krishnan_PLSVD11}. Overall, these methods aim at the identification of weight vectors, from which components are defined as linear transformations of the observed variables. Under each particular method, the weights are chosen so that the corresponding components meet a particular criterion. For example, given a data matrix $\textbf{X}$ containing $n$ observations of a $p$-variate variable $x$ (with $n\geq 1$, $p\geq 1$), the goal of PCA is to identify $r\leq p$ unit weight vectors  that define $r$ mutually orthogonal principal components with maximal variances; the matrix of principal components ${\cal X}= \textbf{X} A^{\top}$ then consists of linear combinations of the $p$ columns of $\textbf{X}$, with the matrix of weights $A$ given by the eigenvectors associated with the $r$ largest eigenvalues of the sample variance matrix $\textbf{X}^{ \top} \textbf{X}$. On the other hand, given two data matrices $\textbf{X}$ and $\textbf{Y}$, that gather the $n\geq 1$ observations for a pair of variables $(x,y)$, with $x\in\R^p$ and $y\in\R^q$, $p,q\geq 1$, the goal of CCA and PLS is to model the relationship between $x$ and $y$ by identifying weight vectors that define components with maximal association. 

Although CCA, which targets components with maximal correlation, is sometimes considered as a PLS technique, the PLS qualifier usually specifically refers to the class of methods that target components with maximal covariance \citep{Wegelin00}. The family of PLS methods consists of a number of techniques, such as PLS Regression, PLS-W2A or PLS-SVD \citep{Wegelin00,overviewPLSRosipal06}. PLS Regression treats the two sets of variables asymmetrically: it focuses on the construction of components from one set of variables, which are then considered as predictors of the second set of variables (the response). On the other hand, PLS-W2A, PLS-SVD and CCA adopt a more symmetrical perspective, and aim at the identification of two sets of weight vectors defining two sets of components. In particular, PLS-SVD, sometimes also referred to as PLS-SB or PLS-C \citep{Wegelin00, Sampson89_PLSSVD, Krishnan_PLSVD11}, is based on the Singular Value Decomposition (SVD) of the sample covariance matrix $\textbf{X}^{\top} \textbf{Y}$, and defines the two sets of weight vectors as left and right singular vectors of  $\textbf{X}^{\top} \textbf{Y}$, respectively. More precisely, the optimization problem is given by
$$\underset{a_1 \in \R^p ,b_1 \in \R^q}{\text{argmax}} \text{ }  a_1^{\top} \textbf{X}^{\top} \textbf{Y} b_1 \quad \text{s.t. } \quad a_1^{\top} a_1 = b_1^{\top} b_1 = 1,$$
and 
\begin{eqnarray*}
   \underset{a_j \in \R^p ,b_j \in \R^q}{\text{argmax}} \text{ } a_j^{\top} \textbf{X}^{\top} \textbf{Y} b_j \quad \text{s.t.} && a_j^{\top} a_j = b_j^{\top} b_j = 1,\\
   && a_j^{\top} \textbf{X}^{\top} \textbf{Y} b_k = 0,\\[0.3cm]
   && \text{for } k, j \in \lbrace 1, \dots, \text{min}(p, q)\rbrace, k \neq j. 
\end{eqnarray*}
\noindent In contrast, CCA, which also treats the two sets of observed variables symmetrically but aims at the identification of two sets of components with maximal correlation, corresponds to the following optimization problem
$$\underset{a_1 \in \R^p ,b_1 \in \R^q}{\text{argmax}} \text{ }  a_1^{\top} \textbf{X}^{\top} \textbf{Y} b_1 \quad \text{s.t. } \quad a_1^{\top} \textbf{X}^{\top} \textbf{X} a_1 = b_1^{\top} \textbf{Y}^{\top} \textbf{Y} b_1 = 1$$
and 
\begin{eqnarray*}
\underset{a_j \in \R^p ,b_j \in \R^q}{\text{argmax}} \text{ } a_j^{\top} \textbf{X}^{\top} \textbf{Y} b_j  \quad \text{s.t. } && a_j^{\top} \textbf{X}^{\top} \textbf{X} a_j = b_j^{\top} \textbf{Y}^{\top} \textbf{Y} b_j = 1,\\
&& a_j^{\top} \textbf{X}^{\top} \textbf{X} a_k = b_j^{\top} \textbf{Y}^{\top} \textbf{Y} b_k = 0,\\[0.3cm]
&& \text{for } k, j \in \lbrace 1, \dots, \text{min}(p, q)\rbrace, k \neq j.
\end{eqnarray*}
For the sake of completeness, we shall further recall that both PLS Regression and PLS-W2A are iterative methods, based on a principle called deflation, which is applied iteratively to guarantee some particular orthogonality properties \citep{Wegelin00, Wold85, overviewPLSRosipal06, PLSR_hoskuldsson88}.

Over the last two decades, several probabilistic formulations of these various dimension-reduction techniques have been introduced, first under a Gaussian setting. They include the Probabilistic PCA (PPCA) \citep{PPCA_Tipping99}, the Probabilistic CCA (PCCA) \citep{PCCA_Bach05}, as well as several versions of Probabilistic PLS (PPLS) \citep{PPLSR_Li15, PPLSR_Zheng16, elBouhaddani17}. Regarding these three probabilistic formulations of the PLS, both \cite{PPLSR_Zheng16} and \cite{PPLSR_Li15} focus on PLS Regression, while \cite{elBouhaddani17} consider a symmetrical PLS approach. Overall, all these probabilistic formulations rely on structural equations that define the observed variables as linear combinations of some latent variables plus some Gaussian noise. Parameter estimation under these latent variable models is then usually performed via an Expectation-Maximization (EM) algorithm \citep{Dempster77_EM}, under appropriate constraints imposed on the model parameters to mimic their non-probabilistic counterpart. Giving access to all the likelihood-based inference machinery, these probabilistic formulations have a number of advantages compared to their standard formulation counterpart \citep{overviewPLSRosipal06, Smilde04}. The estimation can deal with missing data, while still being computationally efficient \citep{PPCA_Tipping99, PPLSR_Zheng16}. Moreover, covariates can be included in the model \citep{Poisson_PPCA_Chiquet17}, and penalized versions of the likelihood can be used to encourage sparsity or structured sparsity, {\it e.g.}, in a high-dimensional framework \citep{Guan09_SPPCA, Zeng17_SPPCA, Park17_SPPCA}. Finally, the probabilistic formulation is very versatile, and turns several complex settings into natural extensions of the simple Gaussian ones mentioned above. For example, probabilistic PCA models have been proposed for binary data and count data \citep{Probabilistic_Count_Durif19, Poisson_PPCA_Chiquet17}. Extensions to even more complex settings, including mediation analysis where three sets of observed variables are involved, have also been proposed \citep{Derkach_LVMA2019}.

To recap, probabilistic formulations of dimension-reduction techniques enjoy a number of appealing properties. However, appearances can be deceptive, and we will show in this article that some caution is needed when developing and applying them. Indeed, despite their apparent ability to fully capture the relationships among the variables under study, some of these models manage to do so under very particular distributions only, which greatly limits their applicability and interest. In particular, they are most often misspecified in practice and, when they are not, their parameters could be estimated under much simpler models. For illustration, we will mainly focus on the probabilistic PLS model proposed by \cite{elBouhaddani17}, which we will simply refer to as the PPLS model from now on. In Section \ref{Sec:orginal_PPLS_model}, we recall the principle of the PPLS model as proposed by \cite{elBouhaddani17}, and emphasize that it can be regarded as a probabilistic formulation of PLS-SVD. In Section \ref{Sec:Limitations} we show that this PPLS model suffers from the aforementioned defect, and actually defines a set of very particular distributions for $(x,y)$, under which components of maximal covariance are also of respective maximal variances. We propose an alternative probabilistic formulation of PLS-SVD in Section \ref{Section:Alternative_model}. In Section \ref{Section:link_newPPLS_PCCA}, we briefly discuss the connection of the proposed model with the probabilistic formulation of the CCA proposed by \cite{PCCA_Bach05}. We turn our attention to the probabilistic formulation of PLS Regression proposed by \cite{PPLSR_Li15} in Section \ref{Section:PPLSR_model}, which we will refer to as the PPLSR model, and show that it suffers from similar limitations.  In Section \ref{Sec:simul}, we present numerical examples to illustrate the limitations of the original PPLS model of \cite{elBouhaddani17}. Concluding remarks are presented in Section \ref{Section:Discussion}.

\section{PPLS models}\label{Section:PPLS_model}
\subsection{The  PPLS model proposed by \cite{elBouhaddani17}}\label{Sec:orginal_PPLS_model}

\begin{figure}[h]
\centering
\begin{minipage}[c]{0.58\linewidth}
\begin{center}
\begin{tikzpicture}[scale = 0.53, auto,swap]
\node[var] (X1)at(8,0.4){$x$};
\node[unobsvar] (T1)at(9.8,2){$t$};
\node[var] (M1)at(14.4,0.4){$y$};
\node[unobsvar] (U1)at(12.6,2){$u$};
\node[unobsvar] (E)at(6.2,2){$e$};
\node[unobsvar] (F)at(16.2,2){$f$};
\draw[edge] (T1)--(X1);
\draw[edge] (T1)--(U1);
\draw[edge] (U1)--(M1);
\draw[edge] (F)--(M1);
\draw[edge] (E)--(X1);
\end{tikzpicture}
\end{center}
\end{minipage}\hfill
\begin{minipage}[c]{0.42\linewidth}
\begin{center}
\begin{tikzpicture}[scale = 0.53, auto,swap]
\node[var] (X1)at(8,0.4){$x$};
\node[unobsvar] (T1)at(9.8,2){$t$};
\node[var] (M1)at(11.6,0.4){$y$};
\node[unobsvar] (E)at(6.2,2){$e$};
\node[unobsvar] (F)at(13.4,2){$f$};
\draw[edge] (T1)--(X1);
\draw[edge] (T1)--(M1);
\draw[edge] (F)--(M1);
\draw[edge] (E)--(X1);
\end{tikzpicture}
\end{center}
\end{minipage}

\begin{minipage}[c]{0.6\linewidth}
\begin{center}
\textbf{A}
\end{center}
\end{minipage}\hfill
\begin{minipage}[c]{0.4\linewidth}
\begin{center}
\textbf{B}
\end{center}
\end{minipage}
\begin{flushleft}
\caption{Graphical representations for: \textbf{A} - The PPLS model proposed by \cite{elBouhaddani17} and recalled in Equation \eqref{Eq:initial_PPLS_model}. \textbf{B} - Our PPLS-SVD model given in Equation \eqref{Eq:model2}. Note that the later has one set of latent variables $t$ only. Moreover, the structure of the noise parts $e$ and $f$ differs between the two models (see Equations \eqref{Eq:initial_PPLS_model} and \eqref{Eq:model2} below). In both models, $x$ and $y$ are the observed variables whereas circled variables are unobserved. }\label{Fig:DAGS}
\end{flushleft}
\end{figure}

The PPLS model proposed by \cite{elBouhaddani17} can be graphically represented as depicted in Figure \ref{Fig:DAGS} \textbf{A}. More specifically, it is defined by the following structural equations, which relate the two observed sets of variables $x\in \R^p$ and $y\in\R^q$ to two sets of latent variables $t\in \R^r$ and $u\in \R^r$, with $r <  \min(p,q)$, 
\begin{equation}
x = tW^{\top} + e, \quad y = uC^{\top} + f, \quad u = t B + h, \label{Eq:initial_PPLS_model}
\end{equation}
together with the following constraints \textbf{(a)}-\textbf{(j)} on the model parameters

\begin{itemize}
\item[(a)] $t \sim \mathcal{N}(0_r, \Sigma_t)$.
\item[(b)] $\Sigma_t$ is a $r \times r$ diagonal matrix, with strictly positive diagonal elements. 
\item[(c)] $e \sim \mathcal{N}(0_p, \sigma_e^2 I_p)$. \quad \quad  \quad \textbf{(d)} $f \sim \mathcal{N}(0_q, \sigma_f^2 I_q)$. \quad \quad \quad \textbf{(e)} $h \sim \mathcal{N}(0_r, \sigma_h^2 I_r)$.
\item[(f)] $e$, $f$ and $h$ are independent.
\item[(g)] $W$ and $C$ are respectively $p \times r$ and $q \times r$ semi-orthogonal matrices.
\item[(h)] $B$ is a diagonal matrix, with strictly positive diagonal elements.
\item[(i)] the diagonal elements of $\Sigma_t B$ are strictly decreasingly ordered.
\item[(j)] $r < \text{min}(p,q)$.
\end{itemize}
Here $I_p$ denote the identity matrix of size $p \times p$, and $0_p$ the vector $(0, \dots,0)$ of size $p$. The parameters of the model are given by $\theta = (W, C, B ,\Sigma_t, \sigma_e^2, \sigma_f^2, \sigma_h^2 )$. 
In particular, matrices $W = (W_1, \cdots, W_r)$ and $C=(C_1, \cdots, C_r)$ contain the two sets of weight vectors; note that they are the ``true weights'', defined from the theoretical distribution of ($x$, $y$). Given estimates $\widehat W$ and $\widehat C$ of these quantities, two sets of empirical components can be defined as linear combination of the two sets observed variables. In this work, we will mostly focus on components defined as $\widehat{\cal X} = \textbf{X} \widehat W$ and $\widehat{\cal Y} = \textbf{Y} \widehat C$; we recall that, when working with latent variable models, an alternative strategy consists in using appropriate conditional expectations of the latent variables; see \cite{PCCA_Bach05} and Section \ref{Section:Alternative_model} below for more details. We shall further stress that in either case, the components do not directly correspond to the latent variables $t$ and $u$. In particular, ${\cal x} = x W = t + eW$ and ${\cal y} = y C = u + fC$ typically differ from $t$ and $u$, respectively.

Under the constraints \textbf{(a)}-\textbf{(j)}, \cite{elBouhaddani17} establish the identifiability of their model (up to sign for the columns of parameters $W$ and $C$). In particular, the identifiability of parameters $W$ and $C$ is given by the following Proposition.
\begin{Proposition}\label{prop1}
Under the PPLS model given in Equation \eqref{Eq:initial_PPLS_model} along with the constraints $\rm{\textbf{(a)}-\textbf{(j)}}$, the columns of $W$ and $C$ are the uniquely defined (up to sign) left and right singular vectors corresponding to the $r$ largest singular values of ${\rm Cov}(x,y)$, respectively.
\end{Proposition}
This result has already been established by \cite{elBouhaddani17} in their Lemma 1, so we here only recall the sketch of the proof. Under the PPLS model, we have ${\rm Cov}(x,y) = W \Sigma_t B C^{\top}$, where $W$ and $C$ are semi-orthogonal matrices, and $\Sigma_t B$ is diagonal with strictly positive decreasingly ordered diagonal elements. It follows that the first $r$ non-null singular values of ${\rm Cov}(x,y)$ are all distinct, and are the diagonal elements of $\Sigma_t B$. As a result, the columns of $W$ and $C$ are uniquely defined (up to sign) as the first $r$ left and right singular vectors of ${\rm Cov}(x,y)$, respectively. 

Although they do not mention it, their model can therefore be regarded as a probabilistic formulation of PLS-SVD. In particular, this means that the two sets of components ${\cal x} = x W$ and ${\cal y} = y C$ coincide with the two sets of components with maximal covariance, targeted by the PLS-SVD.

However, we establish in Section \ref{Sec:Limitations} that the two sets of weights $W$ and $C$, which are the theoretical solutions of the PPLS model, are also necessarily theoretical solutions of two PPCA models for $x$ and $y$, respectively. In other words, the PPLS model defines a set of very particular distributions for $(x,y)$ under which the two sets of components with maximal covariance, ${\cal x} = x W$ and ${\cal y} = y C$, are also necessarily of respective maximal variances.

\subsection{Limitation of the PPLS model proposed by \cite{elBouhaddani17}}\label{Sec:Limitations}

In Supporting Information A, we establish the following result.

\begin{Proposition}\label{Prop:limitationPPLS}
Under the PPLS model given in Equation \eqref{Eq:initial_PPLS_model} along with the constraints $\rm{\textbf{(a)}}$-$\rm{\textbf{(j)}}$, the columns of $W$ and $C$ are eigenvectors corresponding to the $r$ largest eigenvalues of ${\rm Var}(x)$ and ${\rm Var}(y)$, respectively.
\end{Proposition}

Proposition \ref{Prop:limitationPPLS} notably implies that, under the PPLS model, the two sets of components ${\cal x} = x W$ and ${\cal y} = y C$ are not only of maximal covariance (as implied by Proposition \ref{prop1}), but they are also necessarily of respective maximal variances. This result, whose proof is given in Supporting Information A, can equivalently be deduced from the fact that solutions $W$ and $C$ of the PPLS model are also necessarily solutions of two PPCA models, for $x$ and $y$ respectively. More precisely, the PPLS model implies that both $x$ and $y$ fulfill 
the following PPCA model, presented here for a generic observed variable $z \in \mathbb{R}^d$
\begin{equation}
    z= v  V^{ \top} + g, \label{Eq:PPCA}
\end{equation}
under the constraints
\begin{itemize}
\item[(\boldmath{$\alpha$})] $v \sim \mathcal{N}(0_r, \Sigma_v)$. \quad \quad \quad \quad \textbf{(\boldmath{$\beta$})} $\Sigma_v$ is a $r \times r$ diagonal matrix, with strictly positive diagonal elements. 
\item[(\boldmath{$\gamma$})] $g \sim \mathcal{N}(0_d, \sigma_g^2 I_d)$. \quad \quad  \quad \textbf{(\boldmath{$\delta$})} $V$ is a $d \times r$ semi-orthogonal matrix.
\item[(\boldmath{$\epsilon$})] $r <d$.
\end{itemize}
This PPCA model is a variation of the one introduced by \cite{PPCA_Tipping99}; see Supporting Information B for more details. 

First consider this PPCA model for the observed variable $x\in\R^p$. By comparing, on the one hand, constraints \textbf{(a)}, \textbf{(b)}, \textbf{(c)}, \textbf{(g)} and \textbf{(j)} with constraints \textbf{(\boldmath{$\alpha$})}-\textbf{(\boldmath{$\epsilon$})}, and, on the other hand, Equation \eqref{Eq:PPCA} and the first equation in Equation \eqref{Eq:initial_PPLS_model}, it appears that the unique solution $W$ of the PPLS model necessarily corresponds to one of the possibly many solutions $V$ of this PPCA model for $x$. More precisely, when the solution of the PPCA model for $x$ is unique (up to sign), that is when the diagonal elements of $\Sigma_t$ are all distinct, then the $r$ largest eigenvalues of ${\rm Var}(x)$ are all of algebraic multiplicity equal to one, the associated eigenvectors are uniquely defined (up to sign), and they correspond to the columns of $V$. They are also the columns $W_1, \ldots, W_r$ of $W$, although not necessarily in the same order; columns of $W$ and $V$ are in the same order if, and only if, the diagonal elements of $\Sigma_t$ are in 
decreasing order too. Now, if the diagonal elements of $\Sigma_t$ are not all distinct, then the solution $V$ of the PPCA model for $x$ is not unique, but the columns of $W$ still necessarily constitute one of these solutions, that is one particular set of eigenvectors corresponding to the $r$ largest eigenvalues of ${\rm Var}(x)$. 

Similarly, the PPLS model implies that the PPCA model above holds for the observed variable $y\in\R^q$ too, and that the unique solution $C$ of the PPLS model necessarily corresponds to one of the possible solutions of this PPCA model for $y$. More precisely, if the diagonal elements of $\Sigma_tB^2$ are all distinct, then the columns of $C$ correspond to the uniquely defined $r$ eigenvectors associated with the $r$ largest eigenvalues of ${\rm Var}(y)$. On the other hand, if the diagonal elements of $\Sigma_t B^2$ are not all distinct, then the columns of $C$ still constitute one of the solutions of the PPCA model for $y$; in particular, they are one of the possible sets of eigenvectors for the $r$ largest eigenvalues of ${\rm Var}(y)$.

Putting all this together, the PPLS model of \cite{elBouhaddani17} corresponds to a model where two PPCA models, one for $x$ and one for $y$, are related to each other via the third equation in Equation \eqref{Eq:initial_PPLS_model}. Therefore, the weight matrices $W$ and $C$, solutions of their PPLS model, are also necessarily solutions of two PPCA models for $x$ and $y$, so that their model defines a subset of very particular distributions for $(x,y)$, under which components ${\cal x}= xW$ and ${\cal y}=yC$ are not only of maximal covariance, but also of respective maximal variances. In particular, if the diagonal elements of $\Sigma_t$ are all distinct, and if the same holds true for $\Sigma_t B^2$, the ``solutions'' of the two distinct PPCA models are uniquely defined, and then each of the two marginal distributions of $x$ and $y$ are sufficient to respectively identify each of the two sets of weights that define components with maximal covariance. As will be confirmed in Section \ref{Sec:simul}, this greatly limits the applicability of the PPLS model: it is most often misspecified in practice, and when it is not, two PCAs (or PPCAs) are often sufficient to estimate the weight matrices $W$ and $C$.

\subsection{Alternative probabilistic formulation of the PLS-SVD}\label{Section:Alternative_model}

We now present an alternative probabilistic formulation of the PLS-SVD, named PPLS-SVD, which corrects the main defect of the model proposed by \cite{elBouhaddani17} and defines a broader set of distributions for $(x,y)$. Our overall objective was to keep the same general form as that of \cite{elBouhaddani17}, but with weaker constraints, in such a way that the weights $W$ and $C$ cannot generally be identified from the marginal distributions of $x$ and $y$ only.

In the PPLS model, assumptions \textbf{(a)}-\textbf{(j)} are related to various aspects of the model: the distributions of the errors terms, the distributions of the latent variables, as well as ``direct'' constraints on the model parameter $\theta = (W, C, B ,\Sigma_t, \sigma_e^2,$ $\sigma_f^2, \sigma_h^2 )$. In order to keep the link with the PLS-SVD, we still assume that the weights matrices $W$ and $C$ are semi-orthogonal, and that the variance matrices of the latent variables are diagonal. As a start, we thus only relax the constraints \textbf{(c)} and \textbf{(d)} on the isotropy of the variance matrices for the error terms $e$ and $f$. 
To be as general as possible, we simply assume that these variance matrices are positive semi-definite, so that the error terms $e$ and $f$ are simply two non-degenerate Gaussian vectors. We therefore replace constraints \textbf{(c)} and \textbf{(d)} by constraints \textbf{(c*)} and \textbf{(d*)} presented below. But then, to preserve the identifiability of the model (see below), we have to consider a model with only one set of latent variables, in the same vein as the PCCA model of \cite{PCCA_Bach05}. The PPLS-SVD model, depicted in Figure \ref{Fig:DAGS} \textbf{B}, is then defined by the following two structural equations 
\begin{equation}
x = tW^{\top} + e, \quad y = tC^{\top} + f, \label{Eq:model2}
\end{equation}
under the constraints \textbf{(a)}, \textbf{(b)}, \textbf{(g)}, \textbf{(j)} and
\begin{itemize}
\item[(c*)] $e \sim \mathcal{N}(0_p, \Psi_e)$, with $\Psi_e$ a $p\times p$ semi-positive definite matrix.
\item[(d*)] $f \sim \mathcal{N}(0_q, \Psi_f)$, with $\Psi_f$ a $q\times q$ semi-positive definite matrix.
\item[(f*)] $e$ and $f$ are independent.
\item[(i*)] the diagonal elements of $\Sigma_t$ are strictly decreasingly ordered.
\end{itemize}
Conditions \textbf{(f*)} and \textbf{(i*)} are the analogues of conditions \textbf{(f)} and \textbf{(i)}, respectively, in the case where only one set of latent variables is considered. Further observe that ${\rm Cov}(x,y) = W \Sigma_t C^{\top}$, ${\rm Var}(x) = W \Sigma_t W^{\top} + \Psi_e$, and  ${\rm Var}(y) = C \Sigma_t C^{\top} + \Psi_f$, where $\theta = (W, C, \Sigma_t, \Psi_e, \Psi_f)$ are the parameters of our model. Again, the graphical model presented in Figure \ref{Fig:DAGS} \textbf{B} does not completely specify our model, which is defined through the structural equations in \eqref{Eq:model2}, together with the constraints \textbf{(a)}, \textbf{(b)}, \textbf{(c*)}, \textbf{(d*)}, \textbf{(f*)}, \textbf{(g)}, \textbf{(i*)}, and \textbf{(j)}.

We now present the sketch of the proof of the identifiability of the PPLS-SVD model, which is an adaptation of the one developed by \cite{elBouhaddani17}; we refer to Supporting Information C for a more detailed proof. Consider two pairs of random variables, $(x,y)$ and $(\widetilde{x}, \widetilde{y})$, drawn from two PPLS-SVD models, with respective parameters $\theta = (W, C, \Sigma_t, \Psi_e, \Psi_f)$ and $\widetilde \theta = (\widetilde W, \widetilde C, \widetilde \Sigma_t, \widetilde \Psi_e, \widetilde \Psi_f)$, and respective variance-covariance matrices $\Sigma$ and $\widetilde \Sigma$. Now, assume that $\Sigma = \widetilde \Sigma$. This is equivalent to

\begin{eqnarray}
W \Sigma_t W^{\top} + \Psi_e &=&  \widetilde W \widetilde{\Sigma}_t \widetilde W^{\top} + \widetilde {\Psi}_e, \label{Eq:model2_varX} \\
C \Sigma_t  C^{\top} + \Psi_f &=& \widetilde C \widetilde{\Sigma}_t \widetilde C^{\top} + \widetilde {\Psi}_f, \label{Eq:model2_varY}\\
W \Sigma_t C^{\top} &=& \widetilde W \widetilde{\Sigma}_t \widetilde C^{\top} . \label{Eq:model2_cov}
\end{eqnarray}
Matrices $W$, $C$, $\widetilde W$, and $\widetilde C$ are all semi-orthogonal, and both $\Sigma_t$ and $\widetilde{\Sigma}_t$ are diagonal with strictly decreasing diagonal elements. 
As detailed in Supporting Information C, Equation \eqref{Eq:model2_cov} implies that $ \Sigma_t =  \widetilde{\Sigma}_t $, $W = \widetilde W J$ and $C = \widetilde C J$, with $J $ a diagonal matrix with $\pm 1$ elements on the diagonal. 
Then, Equation \eqref{Eq:model2_varX} implies that ${\Psi_e} = \widetilde {\Psi}_e$, while Equation \eqref{Eq:model2_varY} implies that ${\Psi_f} = \widetilde {\Psi}_f$. As a result, the parameters of the PPLS-SVD model given in Equation \eqref{Eq:model2} are identifiable (up to sign for the columns of $W$ and $C$). In particular, because ${\rm Cov}(x,y) = W \Sigma_t C^{\top}$, parameters $W$ and $C$ are identified (up to sign) as the first $r$ left and right singular vectors of ${\rm Cov}(x,y)$, respectively. 

Moreover, because ${\rm Var}(x) = W \Sigma_t W^{\top} + \Psi_e$, and  ${\rm Var}(y) = C \Sigma_t C^{\top} + \Psi_f$, with $\Psi_e$ and $\Psi_f$ two positive semi-definite matrices, we shall stress that $W$ and $C$ can generally not be identified from the eigendecomposition of ${\rm Var}(x)$ and ${\rm Var}(y)$, respectively. In other words, the two sets of weight matrices $W$ and $C$ define components with maximal covariance, which are not necessarily of respective maximal variances, and $W$ and $C$ cannot generally be identified separately from the marginal distributions of $x$ and $y$. 
Our PPLS-SVD model can therefore be regarded as a more general probabilistic formulation of the PLS-SVD, which defines a much broader and interesting set of distributions than the original PPLS model of \cite{elBouhaddani17}.

We will now conclude this Section with a few remarks on our model. First, we shall stress that the residuals, $e$ and $f$ of our model, may be more than simple noise terms. Indeed, they consist of everything that is not in the shared part between $x$ and $y$. In particular, $e$ may contain some signal from additional latent variables specific to $x$, as in the probabilistic PLS Regression model proposed by \cite{PPLSR_Zheng16}; see Equation \eqref{Eq:initial_PPLSR_model2} below. Similarly, $f$ may contain some signal from additional latent variables specific to $y$.

Second, two sets of components can be defined as linear transformations of $x$ and $y$, respectively.  As above, just as under the standard PLS-SVD \citep{Wegelin00}, a first strategy consists in defining ${\cal x} = x W$ and ${\cal y} = y C$. Following \cite{PCCA_Bach05}, alternative components are defined as $\mathcal{x}^{*} = {\rm E}(t \vert x ; \theta)$ and $\mathcal{y}^{*} = {\rm E}(t \vert y ; \theta)$. As ${\rm E}(t \vert x ; \theta) = x \big( W \Sigma_t W^{ \top} + \Psi_e \big)^{-1} W \Sigma_t$ and ${\rm E}(t \vert y; \theta) = y \big( C \Sigma_t C^{ \top} + \Psi_f \big)^{-1} C\Sigma_t$, ${\cal x^*}$ and ${\cal y*}$ are linear transformations of $x$ and $y$ too, but the linear sub-spaces corresponding to ${\cal x}^*$ and ${\cal y}^*$, and  ${\cal x}$ and ${\cal y}$, respectively, differ, unless $\Psi_e$ and $\Psi_f$ are zero matrices \citep{PCCA_Bach05}.

Finally, a last remark concerns the estimation of the parameters under our model. Parameters $W$ and $C$ could be estimated by performing a simple SVD of the covariance matrix ${\rm Cov}(\textbf{X}, \textbf{Y})$. Alternatively, an EM algorithm would yield estimates for all the parameters $\theta$, while taking into account all the constraints of the model. It would further allow various extensions, such as the inclusion of covariates, etc. However, the derivation of the EM algorithm is less straightforward under our extended model than under the original PPLS model. In particular, the updates in each of the M-steps of the EM for the parameters $W$ and $C$ require an optimization problem over the Stiefel Manifold to be solved \citep{Stiefel_Wen2010, Stiefel_Siegel2019}, while these updates have closed-form expressions under the original PPLS model of \cite{elBouhaddani17}. Although we have not fully devised it, additional details on a possible EM algorithm are presented in Supporting Information D.

\subsection{Link between the alternative probabilistic formulation of the PLS-SVD presented in Section 2.3 and the probabilistic formulation of the CCA proposed by \cite{PCCA_Bach05}}\label{Section:link_newPPLS_PCCA}

As mentioned in the Introduction, \cite{PCCA_Bach05} proposed a probabilistic formulation of the CCA. Their PCCA model is defined by the following structural equations, which relate the two sets of observed variables $x\in \R^p$ and $y\in\R^q$ to one set of latent variables $t\in \R^r$, 
\begin{equation*}
x = t \widetilde W^{\top} + e, \quad y = t \widetilde C^{\top} + f, \label{Eq:modelCCA}
\end{equation*}
together with the constraints
\begin{itemize}
\item[(\boldmath{$\tilde \alpha$})] $t \sim \mathcal{N}(0_r, \Sigma_{t})$.
\item[(\boldmath{$\tilde \beta$})] $\Sigma_{t}$ is the $r \times r$ identity matrix.
\item[(\boldmath{$\tilde \gamma$})] $e \sim \mathcal{N}(0_p, \Psi_e)$, with $\Psi_e$ a $p\times p$ semi-positive definite matrix.
\item[(\boldmath{$\tilde \delta$})] $f \sim \mathcal{N}(0_q, \Psi_f)$, with $\Psi_f$ a $q\times q$ semi-positive definite matrix.
\item[(\boldmath{$\tilde \epsilon$})] $e$ and $f$ are independent.
\item[(\boldmath{$\tilde \zeta$})] $r < \text{min}(p,q)$.
\end{itemize}

Both the PPCA model and our PPLS-SVD model proposed in Section \ref{Section:Alternative_model} can be graphically represented as in Figure \ref{Fig:DAGS} \textbf{B}, and they share the same structural equations (see Equation \eqref{Eq:model2}). However, as they are probabilistic formulations of two different dimension-reduction techniques, they differ in the constraints imposed on their parameters. More precisely, constraint \textbf{(b)} in Section \ref{Section:Alternative_model} differs from constraint \textbf{(\boldmath{$\tilde \beta$})} above, and the PPLS-SVD model further imposes constraint \textbf{(g)} on parameters $W$ and $C$. In other words, in our probabilistic formulation of the PLS-SVD, if we no longer require $W$ and $C$ to be semi-orthogonal matrices, and if we further require $\Sigma_t$ to be the identify matrix, the model will coincide with the PCCA of \cite{PCCA_Bach05}.

Note, the number of degrees of freedom for the variance matrices of the noise parts is the same in the two models. However, the number of independent parameters "of interest" under the PCCA of \cite{PCCA_Bach05} is $(p+q) r$, as parameters $\widetilde W$ and $\widetilde C$ have $p  r$ and $q r$ degrees of freedom, respectively, while it equals $(p+q-r)r$ under our PPLS-SVD, as  parameter $\Sigma_t$ has $r$ degrees of freedom and parameters $W$ and $C$ have $p r - \dfrac{r (r+1)}{2}$ and $q  r - \dfrac{r (r+1)}{2}$ degrees of freedom, respectively. In other words, our PPLS-SVD model is more constrained than the PCCA model proposed by \cite{PCCA_Bach05}.

\subsection{Limitation of the PPLSR model proposed by \cite{PPLSR_Li15} and alternative probabilistic formulation of PLS Regression proposed by \cite{PPLSR_Zheng16}}\label{Section:PPLSR_model}

As mentioned in the Introduction, probabilistic formulations of other PLS methods have been proposed in the literature. For example, \cite{PPLSR_Li15} propose a probabilistic formulation of PLS Regression. Their PPLSR model is defined by the following structural equations, which relate the two sets of observed variables $x\in \R^p$ and $y\in\R^q$ to one set of latent variables $t\in \R^r$, 
\begin{equation*}
x = tW^{\top} + e, \quad y = tC^{\top} + f, \label{Eq:initial_PPLSR_model}
\end{equation*}
under the constraints
\begin{itemize}
\item[(a)] $t \sim \mathcal{N}(0_r, I_r)$. \quad \quad \quad \textbf{(b)} $e \sim \mathcal{N}(0_p, \sigma_e^2 I_p)$. \quad \quad \quad \textbf{(c)} $f \sim \mathcal{N}(0_q, \sigma_f^2 I_q)$.
\item[(d)] $e$ and $f$ are independent.
\end{itemize}
Although they do not clearly state it, \cite{PPLSR_Li15} implicitly assume that $r < p$, as $t$ is a ``low-dimensional representation'' of $x$. Contrary to our PPLS-SVD model given in Equation \eqref{Eq:model2} in Section \ref{Section:Alternative_model}, the weight matrices $W$ and $C$ are not supposed to be semi-orthogonal, and the components of the latent variable $t$ are not only independent of each other, but also of unit variance. In that sense, the PPLSR model is inspired by the factor analysis model \citep{FactorA1994}, just as the PPCA model proposed by \cite{PPCA_Tipping99} and recalled in Supporting Information B. More importantly, the error terms $e$ and $f$ are assumed to be of isotropic variances, just as in the original PPLS model proposed by \cite{elBouhaddani17}. \cite{PPLSR_Li15} do not study the identifiability of their model, but, if $r < p$, it is easy to show that parameters $\sigma_e^2$ and $\sigma_f^2$ are identifiable, and parameters $W$ and $C$ are identifiable up to an orthogonal transformation.

As a matter of fact, if $r < \text{min}(p,q)$, the PPLSR model corresponds to the PCCA model proposed by \cite{PCCA_Bach05}, but with the additional constraint of isotropic variances for the error terms. Just as in the original PPLS model of \cite{elBouhaddani17}, this constraint is too strong: if $r < \text{min}(p,q)$, the PPLSR model of \cite{PPLSR_Li15} defines a very particular set of distributions, under which the theoretical solutions of the PPLSR (and hence, of the PCCA) model, $W$ and $C$, are also necessarily solutions of two PPCA models (in the sense of \cite{PPCA_Tipping99}) for $x$ and $y$, respectively.

In the context of PLS Regression, it is maybe more sensible to assume that $q \leq r < p$. 
But then, the PPLSR model proposed by \cite{PPLSR_Li15} still defines a set of particular distributions for $(x,y)$, under which the solution $W$ of the PPLSR is also necessarily solution of a PPCA model for $x$. This observation confirms what \cite{PPLSR_Zheng16} already suggested: the PPLSR model proposed by \cite{PPLSR_Li15} is not an appropriate probabilistic formulation of PLS Regression.

\cite{PPLSR_Zheng16} then proposed an extension of the PPLSR model, defined by the following structural equations, which relate the two sets of observed variables $x\in \R^p$ and $y\in\R^q$ to two sets of latent variables $t\in \R^r$ and $u\in \R^s$, 
\begin{equation}
x = tW^{\top} + u Q^{\top} + e, \quad y = tC^{\top} + f, \label{Eq:initial_PPLSR_model2}
\end{equation}
under the constraints
\begin{itemize}
\item[(a*)] $t \sim \mathcal{N}(0_r, I_r)$. \hspace{1.55cm} \textbf{(b*)} $u \sim \mathcal{N}(0_s, I_s)$.
\item[(c*)] $e \sim \mathcal{N}(0_p, \Sigma_e)$. \hspace{1.4cm} \textbf{(d*)} $f \sim \mathcal{N}(0_q, \Sigma_f)$.
\item[(e*)] $\Sigma_e$ is a $p \times p$ diagonal matrix, with strictly positive diagonal elements. 
\item[(f*)] $\Sigma_f$ is a $q \times q$ diagonal matrix, with strictly positive diagonal elements. 
\item[(g*)] $e$ and $f$ are independent.
\end{itemize}
Again, \cite{PPLSR_Zheng16} implicitly assume that $p>\max(r,s)$.

The inclusion of the additional set of latent variables in the model, $u$, which is related to $x$ only, has the same impact in terms of identifiability of the parameter $W$ as the relaxation of the assumption of isotropic variance for the error terms $e$ that we considered in our PPLS-SVD model (see Equation \eqref{Eq:model2}). In particular, the model proposed by \cite{PPLSR_Zheng16} defines a more general set of distributions than the original model of \cite{PPLSR_Li15}, and the weight parameter $W$ can generally not be identified from the marginal distribution of $x$ only.

It is noteworthy that \cite{PPLSR_Zheng16} do not consider the inclusion of an additional set of latent variables related to $y$ only. Consequently, observing that the variances of the error terms $e$ and $f$ are now simply assumed to be diagonal with positive diagonal elements, the marginal distribution of $y$ is still sufficient to identify $C$ (up to an orthogonal transformation) if this matrix satisfies the condition given in the Theorem 5.1 of \cite{anderson_FA56}, and if $r < q$. In the context of PLS regression, this is not necessarily a limitation. Moreover, the inclusion of a third set of latent variables in the model, related to $y$ only, would define a model equivalent to the PCCA model of \cite{PCCA_Bach05}.

\section{Illustration of the limitation of the PPLS model proposed by \cite{elBouhaddani17}}\label{Sec:simul}

Now, we present results from two simulation studies aimed to illustrate the limitations of the PPLS model of \cite{elBouhaddani17} (see Equation \eqref{Eq:initial_PPLS_model}).  More precisely, our objective is to illustrate the behavior of the estimates for $W$ and $C$ returned by the EM algorithm devised by \cite{elBouhaddani17} under the original PPLS model, depending on whether this model is correctly specified or not. For comparison, we further considered estimates returned by the standard (non-probabilistic) PLS-SVD, and the standard PCA (successively applied on the ``$x$ and $y$ parts'' of the data). The PLS-W2A, which is another symmetrical PLS method that we briefly described in the Introduction (see \cite{Wegelin00, Wold85, overviewPLSRosipal06} for more details), was originally considered too. As expected, estimates returned by the PLS-W2A and PLS-SVD methods were very similar under the original PPLS model (because ${\rm Var} (x W)$ and ${\rm Var}(y C) $ are diagonal under the original PPLS model). 
But, as they were very similar in the second simulation study too, we finally decided to omit the presentation of the results from PLS-W2A. 

We set the dimensions of the observed sets of variables $x$ and $y$ to $p = q= 20$, the dimension of the sets of latent variables to $r=3$, and make the sample size vary in $n \in \lbrace 50, 250, 500, 1000, 5000\rbrace$. In the first simulation study, we work under the same setting as that considered by \cite{elBouhaddani17} in their simulation study. More precisely, data $(\textbf{X},\textbf{Y})$ are generated under the original PPLS model, in the particular case where the diagonal elements of both $\Sigma_t$ and $\Sigma_t B^2$ are all distinct. Weight matrices $W$ and $C$ are randomly drawn from the sets of semi-orthogonal matrices of size $p \times r$ and size $q \times r$, respectively, and the diagonal elements of $\Sigma_t$ and $B$ are respectively set to $\sigma_{t_i}^2 = \text{exp}(-(i-1)/5)$ and $b_i  = 1.5 {\rm exp}(3(i-1)/10)$, for $i\in\{1,2,3\}$, just as in \cite{elBouhaddani17}. As for the variances of $e$, $f$ and $h$, they are chosen so that the signal-to-noise ratios are equal to 0.25: $\sigma_e^2 = 0.4$, $\sigma_f^2 = 4$ and $\sigma_h^2 = 5.33$. The main objective of this first study is to empirically confirm that, when the original PPLS model of \cite{elBouhaddani17} is correctly specified, the weights returned by the corresponding EM algorithm are similar to those returned by two PCAs applied to the $x$ and $y$ parts of the data. In the second simulation study, data are generated under a model similar to the original PPLS model, except that $e$ and $f$ are not of isotropic variance anymore; instead $e$ and $f$ are drawn from multivariate Gaussian variables with arbitrary positive semi-definite variance matrices. More precisely, to make sure we work under really misspecified models where solutions of the PLS-SVD differ from solutions of two PCAs, we chose positive-definite matrices ensuring that eigenvectors of matrices ${\rm Var}(x)$ and ${\rm Var}(y)$ were not too close to the left and right singular vectors of ${\rm Cov}(x,y)$ (using a simple acceptance rejection method). The main objective of this second study is to describe how the solutions of the EM algorithm of \cite{elBouhaddani17} behaves when components of maximal covariance are not of respective maximal variances too, that is when the original PPLS model is misspecified. In both studies, the results are computed over 1000 replicates. For the comparisons of weight vectors, we use the cosine similarity, which simply reduces to the dot product in our case since both the true and estimated weight vectors are of unit length. Results from our simulation studies can be replicated using our R scripts available on GitHub.

\begin{figure}
    \centering
    \includegraphics[scale=0.43]{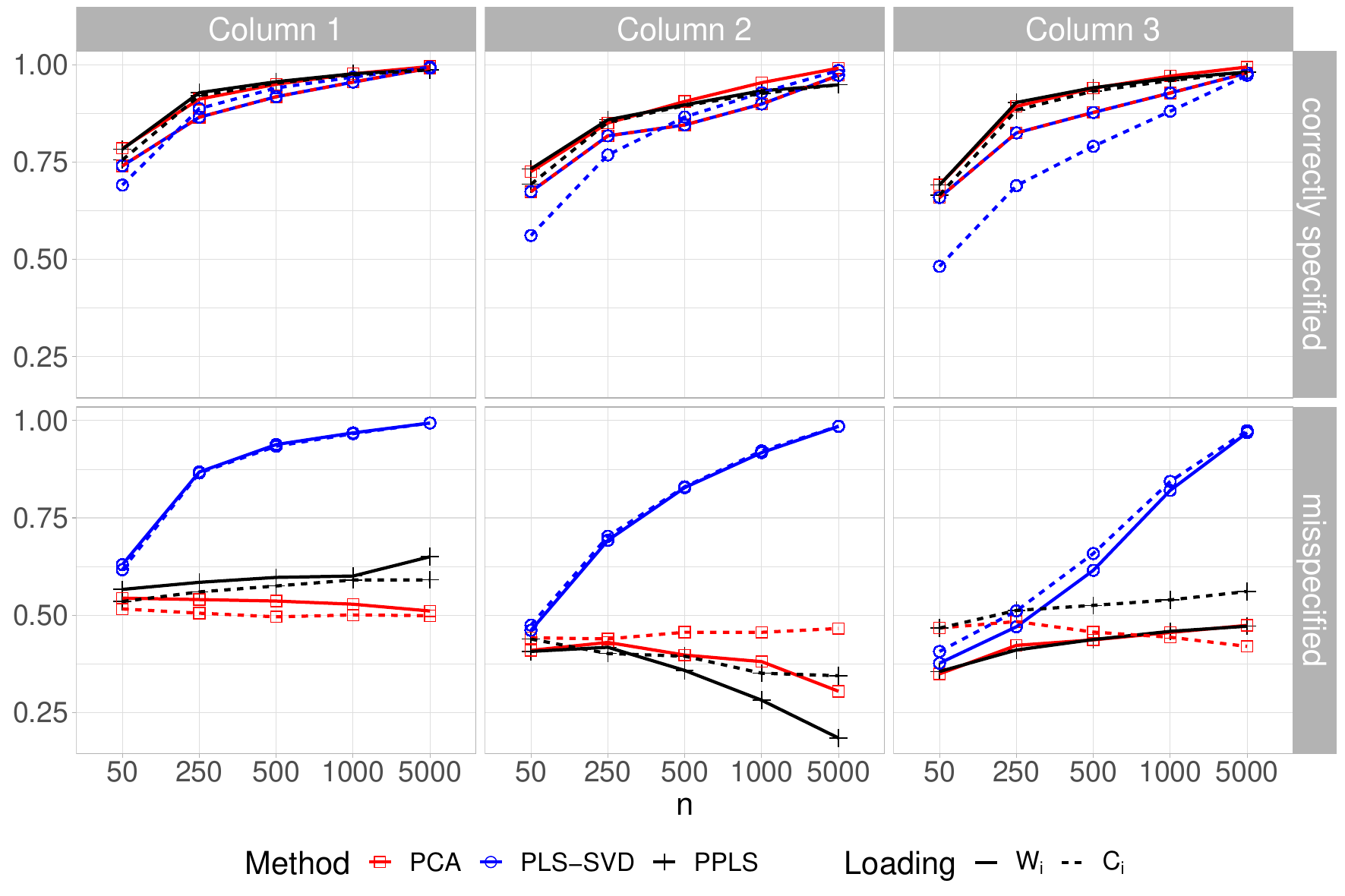}
    \caption{Medians of the cosine similarities (in absolute values) between the true weight vectors and estimates returned by ($i$) the PPLS EM algorithm, ($ii$) two distinct PCAs on $\textbf{X}$ and $\textbf{Y}$, and ($iii$) PLS-SVD on $(\textbf{X},\textbf{Y})$.  The results are computed over 1000 replicates, for $p=q=20$, $r=3$ and different sample sizes $n \in \lbrace 50, 250, 500, 1000, 5000\rbrace$. The top panels correspond to the first simulation study where the original PPLS model is correctly specified, while the bottom panels correspond to the second simulation study where the original PPLS model is misspecified.}\label{Fig:results_simul1}
\end{figure}

Figure \ref{Fig:results_simul1} presents the median of the cosine similarity (in absolute values) between the true weights $W$ and $C$ and their estimates, when the original PPLS model is correctly specified (top panels), and when it is not (bottom panels). Each of the three columns of Figure 1 presents the results for one particular pair $(W_i, C_i)_{i\in\{1,2,3\}}$. We shall stress that  the columns of the estimated weight matrices returned by each of the three compared methods were first re-arranged to make sure they matched the ordering of the true weight matrices, just as in the simulation study conducted by \cite{elBouhaddani17}. 

When the PPLS model is correctly specified (top panels of Figure \ref{Fig:results_simul1}), estimates returned by the EM algorithm under the original PPLS models perform similarly to estimates returned by the other PLS techniques (PLS-SVD and PLS-W2A), and they are all reasonably close to the true weight vectors. In particular, their cosine similarity with the true weight vectors tend to 1 as sample size increases. But, as expected, this is also the case for the estimates returned by two PCAs successively applied to $\textbf{X}$ and $\textbf{Y}$. This empirically confirms that when the diagonal elements of both $\Sigma_t$ and $\Sigma_t B^2$ are all distinct under the original PPLS model, solutions of the PLS-SVD 
coincide with those of the PCAs, limiting the interest of the PLS-SVD in such cases. Moreover, we shall recall that when the diagonal elements of $\Sigma_t$ and/or $\Sigma_t B^2$ are not all distinct, solutions of the PLS-SVD still constitute one of the solutions of the PCAs, indicating that whenever the PPLS model is correctly specified, it is of limited practical interest as its solutions are particular solutions of the PCAs. 

\begin{figure}
    \centering
    \includegraphics[scale=0.43]{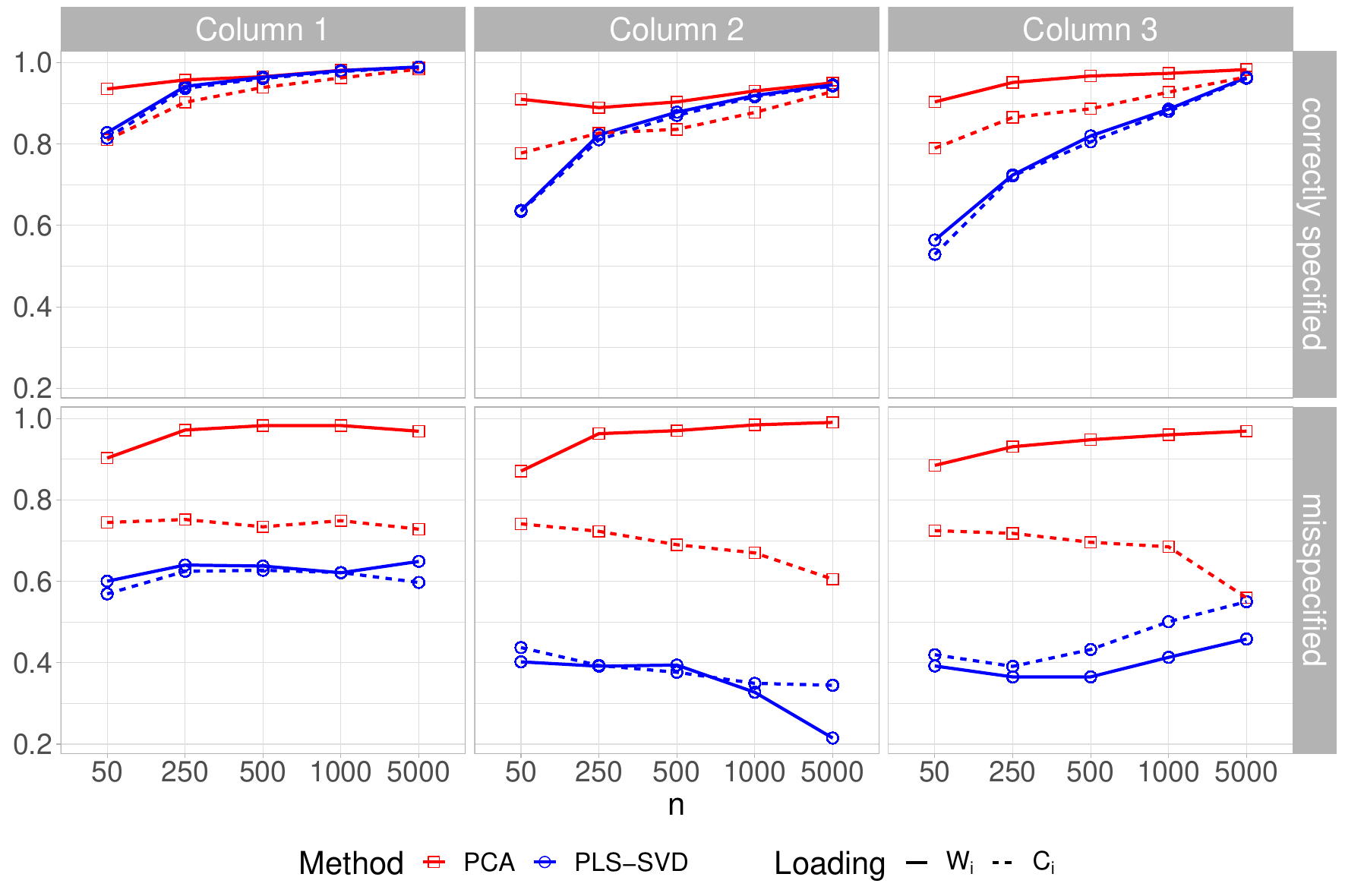} 
    \caption{Medians of the cosine similarities (in absolute values) between the weight vector estimates returned by the EM algorithm devised under the original PPLS model, and those returned by ($i$) two distinct PCAs on $\textbf{X}$ and $\textbf{Y}$, and ($ii$) PLS-SVD on $(\textbf{X},\textbf{Y})$. The results are computed over 1000 replicates, for $p=q=20$, $r=3$ and different sample sizes $n \in \lbrace 50, 250, 500, 1000, 5000\rbrace$. The top panels correspond to the first simulation study where the original PPLS model is correctly specified, while the bottom panels correspond to the second simulation study where it is misspecified.}\label{Fig:results_simul2}
\end{figure}

On the other hand, when the original PPLS model is misspecified (bottom panels of Figure \ref{Fig:results_simul1}), our results show that estimates returned by the two PCAs are quite far from the true weight vectors (as expected, by design), while those returned by the PLS-SVD still perform well. As for the EM algorithm devised under the original PPLS model, it performs much worse than the PLS-SVD, and not much better than the two PCAs. To better describe the estimates returned by the EM algorithm devised under the original PPLS model, Figure \ref{Fig:results_simul2} presents the median of the cosine similarities (in absolute values) between these estimates and those returned by $(i)$ the two distinct PCAs and $(ii)$ the standard PLS-SVD. Interestingly, these results show that, on average, estimates returned by the EM algorithm under the original PPLS model are closer to those returned by the PCAs, especially when the original PPLS model is misspecified.  
Figure 1 in Supporting Information E further presents the box-plots of the absolute value of the cosine similarities between the estimates returned by the EM algorithm devised under the original PPLS model and those returned by $(i)$ two distinct PCAs, and $(ii)$ the standard PLS-SVD, in our second simulation study (when the original PPLS model is misspecified). These box-plots suggest that, when solutions of the PLS-SVD differ from solutions of two PCAs, estimates returned by the EM algorithm proposed by \cite{elBouhaddani17} are generally closer to those returned by the two PCAs. 

All these empirical results confirm that the EM algorithm devised by \cite{elBouhaddani17} under their PPLS model suffers from a severe limitation: in real-life examples, there is no guarantee that the estimated weight vectors it returns really capture the relationship between $x$ and $y$.

\section{Discussion}\label{Section:Discussion}

Our results stress that some caution is needed when developing and applying latent variable models for dimension-reduction: when imposing too strong constraints on the model parameters, a model whose structural equations seem to correctly describe the relationships between the observed variables, may turn out to be too simplistic. It can define very particular distributions, under which parameters of interest could be obtained under simpler models, and this greatly limits its applicability. 
In particular, we showed  that the constraints considered in the probabilistic formulation of PLS-SVD of \cite{elBouhaddani17} are too strong: they imply that the weight matrices $W$ and $C$ of their PPLS model are also necessarily solutions of two distinct PPCA models for $x$ and $y$, respectively. As a result, the original PPLS model defines a very particular subset of distributions for the pair $(x,y)$, under which the two sets of components of maximal covariance are necessarily of respective maximal variances too. Although not striking, this defect severely limits the practical interest of this model. 
 In the same way, the constraints used in the probabilistic formulation of PLS Regression of \cite{PPLSR_Li15} are too strong: they imply that the weight matrix $W$ of their PPLSR model is also necessarily solution of a PPCA model for $x$ (in the sense of \cite{PPCA_Tipping99}). 
 
As shown in the present article, it is sometimes possible to correct for these defects. \cite{PPLSR_Zheng16} proposed an alternative probabilistic formulation of PLS Regression than the one of \cite{PPLSR_Li15}, under which the joint distribution of $(x,y)$ is in general necessary for the identification of the weight matrix allowing the construction of the predictors. On the other hand, in the case of the PPLS model originally proposed by \cite{elBouhaddani17}, we were able to relax some of the constraints, and develop an alternative probabilistic formulation of the PLS-SVD, under which the joint distribution of $(x,y)$ is generally necessary for the identification of the weight matrices. However, the implementation of an EM algorithm for the estimation of the parameters is less straightforward under the PPLS-SVD model. In particular, each M-step of the algorithm requires a numerical optimization step to update the estimates of parameters $W$ and $C$, whereas such updates have closed-form expressions under the original PPLS model. Alternatively, we could propose another version of the model, where parameters $W$ and $C$ would not have to be semi-orthogonal matrices; this would then simplify the EM algorithm. But for the model to be identifiable, we would have to impose $\Sigma_t = I_r$; identifiability would then hold up to an orthogonal transformation for parameters $W$ and $C$. However, in that case, the corresponding model would coincide precisely with the PCCA model proposed by \cite{PCCA_Bach05}; see Section \ref{Section:link_newPPLS_PCCA}. In particular, it would no longer be a probabilistic formulation of the PLS-SVD.

\section*{Acknowledgements} 
The authors are grateful to Anne-Laure Fougères, Thibault Espinasse, Edouard Ollier and Franck Picard for fruitful discussion and comments on earlier versions of this manuscript, and to the referees of Statistica Neerlandica for their valuable suggestions. 

\section*{Disclaimer}
Where authors are identified as personnel of the International Agency for Research on Cancer/World Health Organization, the authors alone are responsible for the views expressed in this article and they do not necessarily represent the decisions, policy or views of the International Agency for Research on Cancer/World Health Organization.

\section*{Data availability statement}

The R code created to generate and analyze the data that support the findings of this study is openly available in repository PPLS-SVD at https://github.com/Etievant/PPLS-SVD.


\section*{Supporting information}

Supporting Information may be found \textcolor{red}{\href{https://onlinelibrary.wiley.com/doi/abs/10.1111/stan.12262}{online}} in the Supporting Information Section at the end of this article.

\end{document}